\documentclass[prd,twocolumn,floatfix,preprintnumbers]{revtex4}

\usepackage{graphicx}
\input{epsf}
\usepackage{epsf}
\usepackage{graphicx,epsfig}
\usepackage{bm}
\usepackage{latexsym,amssymb,amsmath,float}

\newcommand {\ga} {\ {\raise-.5ex\hbox{$\buildrel>\over\sim$}}\ }
\newcommand {\la} {\ {\raise-.5ex\hbox{$\buildrel<\over\sim$}}\ } 
\newcommand{\eqn}[1] {Eq.~(\ref{#1})}

\def\be{\begin{equation}}
\def\ee{\end{equation}}
\def\ba{\begin{eqnarray}}
\def\ea{\end{eqnarray}}
\renewcommand{\(}{\left(} 
\renewcommand{\)}{\right)} 
\renewcommand{\[}{\left[} 
\renewcommand{\]}{\right]}

\begin{document}

\title{Dark Energy from a Phantom Field Near a Local Potential Minimum}
\author{Sourish Dutta and Robert J. Scherrer}
\affiliation{Department of Physics and Astronomy, Vanderbilt University,
Nashville, TN  ~~37235}

\begin{abstract}
We examine dark energy models in which
a phantom field $\phi$ is rolling near a local minimum of its potential $V\(\phi\)$.
We require that $(1/V)(dV/d\phi) \ll 1$, but $(1/V)(d^2 V/d\phi^2)$ can be large.
Using techniques developed in the context of hilltop quintessence, we derive a general
expression for $w$ as a function of the scale factor, and as in the hilltop case,
we find that the dynamics of the field depend on
the value of $\(1/V\)\(d^2 V/d\phi^2\)$ near the mimimum.
Our general result gives a value for $w$ that is within 1\% of the true (numerically-derived)
value for all of the particular cases examined.
Our expression for $w(a)$ reduces to the previously-derived phantom
slow-roll result of Sen and Scherrer in the limit where the potential
is flat, $(1/V)(dV/d\phi) \ll 1$.  
\end{abstract}

\maketitle

\section{Introduction}

Considerable evidence \cite{Knop,Riess1} has accumulated 
suggesting that at least 70\% of the energy density in the
universe is in the form of an exotic, negative-pressure component,
called dark energy.  (See Ref. \cite{Copeland} for a recent
review).

A parameter of considerable importance is the equation of state (EoS) of the dark energy component, defined as the ratio of its pressure to its density:
\be
w=p_{\rm DE}/\rho_{\rm DE}.
\ee
Observations constrain $w$ to be very close to $-1$.
If $w$ is assumed to be constant, then $-1.1 \la w \la -0.9$  \cite{Wood-Vasey,Davis}.
On the other hand, a variety of models have been proposed in which $w$ is time varying.
A common approach is to use a scalar field as the dark energy component.
The class of models in which the scalar field is canonical is dubbed quintessence
\cite{RatraPeebles,TurnerWhite,CaldwellDaveSteinhardt,LiddleScherrer,SteinhardtWangZlatev}
and has been extensively studied.

A related, yet somewhat different approach is phantom dark energy, i.e., a component for which $w<-1$.
The simplest way to realize such a component is to use a scalar field with
a negative kinetic term in its Lagrangian, as first proposed
by Caldwell \cite{Caldwell}. Such models have well-known problems
\cite{CarrollHoffmanTrodden,ClineJeonMoore,BuniyHsu,BuniyHsuMurray},
but nevertheless have been widely studied as potential dark energy candidates
\cite{Caldwell,Guo,ENO,NO,Hao,Aref,Peri,Sami,Faraoni,Chiba,KSS}.

Given the considerable freedom that exists in choosing the potential
function of the scalar field, it is useful to develop general
expressions for the evolution of $w$ which cover a wide range of models.
One such approach is to begin with the observational result that $w$ for dark
energy is very
close to $-1$ today, and to assume that $w$ was always close to $-1$ in
the redshift regime of interest.  With this assumption, one can assume
an expanding background that is very close to $\Lambda$CDM
and solve for the evolution of the scalar field for this case.
This assumption alone is not sufficient to derive a general solution
for the evolution of the scalar field or its equation of state.
However, certain fairly general classes of such models can be solved exactly.
The simplest
such models, explored in Ref. \cite{ScherrerSen1}, assume
a scalar field initially at rest in a potential satisfying the
``slow-roll" conditions:
\ba
\label{SR1}\[\frac{1}{V}\frac{dV}{d\phi}\]^2\ll 1,\\
\label{SR2}\left|\frac{1}{V}\frac{d^2 V}{d\phi^2}\right|\ll 1.
\ea
The first condition insures that $w$ is close to $-1$, while
the two conditions taken together indicate that $(1/V)(dV/d\phi)$ is nearly
constant.
In the terminology of Ref. \cite{CL}, these are ``thawing" models.

For all potentials satisfying these conditions,
it was shown in \cite{ScherrerSen1} that the
behavior of $w$ can be accurately described by
a unique expression depending only on the present-day
values of $\Omega_\phi$ and the initial value of $w$.
In \cite{ScherrerSen2} this result was extended to
phantom models satisfying
Eqs.~(\ref{SR1}-\ref{SR2}), and the $w$ dependence of
these phantom models was shown to be described by
the same expression as in the quintessence case. 

The slow roll conditions, Eqs.~(\ref{SR1}-\ref{SR2}),
while sufficient to ensure $w\simeq-1$ today, are not necessary.
In \cite{DuttaScherrer}, a second possibility was considered,
in which equation (\ref{SR1}) holds, but equation (\ref{SR2}) is relaxed.
This corresponds to a quintessence field rolling near a local
maximum of its potential.  As in the case of slow-roll quintessence,
this case can be solved analytically.  In this case, there is
an extra degree of freedom, the value of $(1/V)(d^2 V/d\phi^2)$,
so that instead of a single solution for the evolution of $w$,
one obtains a family of solutions that
depend on the present-day
values of $\Omega_\phi$ and $w$ and the value of $(1/V)(d^2 V/d\phi^2)$
at the maximum of the potential.  This family of solutions includes
the slow-roll solution as a special case in the limit
where $(1/V)(d^2 V/d\phi^2) \rightarrow 0$.

Here we complete this series of studies by extending
the above result to phantom dark energy models. Since phantom
fields roll up their potentials, the analogous
situation is a phantom rolling close to a local
minimum in its potential. We find that a unique family
of solutions, very similar to the one derived in Ref. \cite{DuttaScherrer}, can be used to approximate the behavior of $w$ in these models. 

\section{Phantom evolution near a minimum}
\label{wexp}
In this section we derive a general expression for the evolution of $w$ for a phantom field near its minimum. Our treatment is similar to \cite{DuttaScherrer}. First consider a minimally coupled phantom field $\phi$ in a potential $V\(\phi\)$. The phantom field has a negative kinetic term in its Lagrangian, leading to an equation of motion
\be
\label{KG}
\ddot{\phi}+3H\dot{\phi}-\frac{dV}{d\phi}=0,
\ee
where $a$ is the scale factor
and $H\equiv\dot{a}/a$ is the expansion rate. Dots
denote derivatives with respect to time and primes
denote derivatives with respect to the field $\phi$.
In a flat universe, the expansion rate is linked to the
total density $\rho_{\rm T}$ via the Friedman equation
(in units where $8\pi G=1$) as
\be
H^2=\rho_{\rm T}/3.
\ee

The evolution of the scale factor is given by:
\be
\frac{\ddot{a}}{a}=-\frac{1}{6}\(\rho_{\rm T}+p_{\rm T}\),
\ee
where $p_{\rm T}$ is the total pressure

Using the transformation
\be
\phi(t)=u(t)/a(t)^{3/2},
\ee
Eq.~(\ref{KG}) becomes 
\be
\label{KGnew}
\ddot{u}+\frac{3}{4}p_{\rm T}u-a^{3/2}V'\(u/a^{3/2}\)=0.
\ee

Now consider a universe consisting of pressureless matter and a
phantom field, where the phantom plays the role of the dark energy.
In order to realistically mimic the observed dark energy,
the phantom must have $w$ close to $-1$ and its energy density
must be roughly constant. The total pressure $p_{\rm T}$ is
then simply given by $p_{\rm T}\approx -\rho_{\phi0}$, where
$\rho_{\phi0}$ is the present day density of the dark energy.
(In what follows, a subscript $0$ always indicates a present day value).
Under this approximation, \eqn{KGnew} becomes:
\be
\label{KGu}
\ddot{u}-\frac{3}{4}\rho_{\phi 0}u-a^{3/2}V'\(u/a^{3/2}\)=0.
\ee
We now apply \eqn{KGu} to a phantom rolling near a local minimum $\phi_{*}$ in its potential. For any $\phi$ close to the minimum, the potential can be expanded in a Taylor series:
\be
V\(\phi\)=V\(\phi_{*}\)+\(1/2\)V''\(\phi_{*}\)\(\phi-\phi_*\)^2+O\(\(\phi-\phi_*\)^3\).
\ee
Substituting the above expansion
into \eqn{KGu}, and taking $V\(\phi_{*}\)=\rho_{\phi 0}$ we obtain 
\be
\label{unew}
\ddot{u}-\[V''\(\phi_{*}\)+\(3/4\)V\(\phi_{*}\)\]u=0.
\ee
With the definition
\be
\label{k}
k\equiv\sqrt{V''\(\phi_{*}\)+\(3/4\)V\(\phi_{*}\)},
\ee
we obtain the general solution to \eqn{unew} to be
\be
u=A\sinh\(kt\)+B\cosh\(kt\).
\ee

The requirement of $w\approx -1$ implies that the potential term dominates the kinetic term in the phantom's energy density. In such a scenario the evolution of the scale factor can be well-approximated by $\Lambda$CDM (see e.g. \cite{gron}):

\be
\label{LCDM a}
a\(t\)=\[\frac{1-\Omega_{\phi 0}}{\Omega_{\phi 0}}\]^{1/3}
\sinh^{2/3}\(t/t_\Lambda\),
\ee
where $\Omega_{\phi 0}$ is the present-day value of the phantom density parameter $\Omega_{\phi}$ and $a=1$ at present. The time $t_{\Lambda}$ is defined as

\be
t_{\Lambda}\equiv 2/\sqrt{3\rho_{\phi 0}}=2/\sqrt{3V\(\phi_{*}\)}.
\ee

These results give
the general solution to equation \eqn{KG} (under the approximations described above) as:
\be
\phi\(t\)=\[\frac{1-\Omega_{\phi 0}}{\Omega_{\phi
0}}\]^{1/2}\frac{A\sinh\(kt\)+B\cosh\(kt\)}{\sinh\(t/t_{\Lambda}\)}, 
\ee
where $A$ and $B$ are arbitrary constants. If we require $\phi\(t=0\)=\phi_{i}$, then $B=0$ and
\be
\label{phishort}
\phi=\frac{\phi_i}{kt_{\Lambda}}\frac{\sinh\(kt\)}{\sinh\(t/t_{\Lambda}\)}.
\ee
This is identical to the evolution of $\phi(t)$ for quintessence near a local
maximum in the potential \cite{DuttaScherrer}.

The EoS parameter $w$ for a phantom is given by
\be
\label{wshort}
1+w=-\frac{\dot{\phi^2}}{\rho_{\phi}}.
\ee
Our requirement that $w\approx-1$ implies that $\rho_\phi\approx\rho_{\phi 0}\approx V\(\phi_{*}\)$.
This, together with \eqn{phishort} and \eqn{wshort}\,
gives an expression for $w(a)$ (normalized to $w_0$,
the present-day value of $w$):
\begin{widetext}
\be
\label{final EOS}
1+w(a)=(1+w_0)a^{-3}
\frac{{\left[ {\sqrt {\Omega _{\phi 0}} kt_\Lambda\cosh
\left[ {k t\left( a \right)} \right] - \sqrt {(1-\Omega_{\phi 0})a^{-3}
+\Omega_{\phi 0}}
\sinh \left[ {k t\left( a \right)} \right]} \right]^2 }}
{{\left[ {\sqrt {\Omega_{\phi 0}}  kt_\Lambda\cosh ( {kt_0})
 - \sinh ( {k t_0})} \right]^2 }},
\ee
\end{widetext}
where $t(a)$ and $t_0$ can be derived from Eq. (\ref{LCDM a}):
\begin{equation}
\label{ta}
t(a) = t_\Lambda \sinh^{-1}\sqrt{\left(\frac{\Omega_{\phi 0} a^3}
{1-\Omega_{\phi 0}}\right)}
\end{equation}
and
\begin{equation}
\label{t0}
t_0 = t_\Lambda \tanh^{-1} \left(\sqrt{\Omega_{\phi 0}}\right).
\end{equation}

For convenience, we now switch to the constant $K\equiv kt_{\Lambda}$. In terms of the phantom potential, $K$ can be written as

\begin{equation}
\label{Kdef}
K = \sqrt{1 + (4/3)V^{\prime \prime}(\phi_*)/V(\phi_*)}.
\end{equation}
indicating that $K$ (which is always $>1$ since $V\(\phi_*\)>0$
and $V''\(\phi_*\)>0$) depends only on the value of the potential and its second derivative at its minimum. In terms of $K$ we can express the evolution of $w$ in the following form:

\begin{widetext}
\begin{equation}
\label{finalfinal}
1 + w(a) = (1+w_0)a^{3(K-1)}\frac{[(F(a)+1)^K(K-F(a))
+(F(a)-1)^K(K+F(a))]^2}
{[(\Omega_{\phi0}^{-1/2}+1)^K(K-\Omega_{\phi0}^{-1/2})
+(\Omega_{\phi0}^{-1/2}-1)^K (K+\Omega_{\phi0}^{-1/2})]^{2}},
\end{equation}
\end{widetext}
where $F(a)$ is given by
\be
F(a) = \sqrt{1+(\Omega_{\phi 0}^{-1}-1)a^{-3}}.
\ee
(Note that $F(a) = 1/\sqrt{\Omega_\phi(a)}$, where $\Omega_\phi(a)$
is the value of $\Omega_\phi$ as a function of redshift, so that
$F(a=1) = \Omega_{\phi0}^{-1/2}$.)

\eqn{finalfinal} is our central result.
We therefore find that, in the limit where
$w$ is slightly less than $-1$
(i.e., where the phantom potential energy dominates its kinetic energy),
all phantom models with a given value of $w_0$ and $\Omega_{\phi 0}$
converge to a unique evolution of $w$ characterised by the value of
$K$ (which depends on $V''\(\phi_*\)/V\(\phi_*\)$, i.e.,
loosely speaking, on the curvature of the potential at its minimum).
For phantom potentials which are very flat, i.e., which satisfy both
slow-roll conditions,
a similar analysis was shown to yield only
a single form of $w(a)$ (Eqn.~(18) of \cite{ScherrerSen2}).
\eqn{finalfinal} captures
the more complex behavior introduced
by the dependence of $w(a)$ on $K$.
It is straightforward to show analytically
that \eqn{finalfinal}
goes over to Eqn.~(18) of \cite{ScherrerSen2} in the
slow-roll limit, $K\rightarrow1$. 

Note that \eqn{finalfinal} has an identical
functional form to the corresponding expression for $w(a)$ for
quintessence models derived in Ref. \cite{DuttaScherrer}.  The only
difference for the phantom models is that $1+w_0 < 0$, producing
a corresponding $w(a) < -1$.  Of course, this is a significant difference
in deriving observational constraints on the models.  Further, there is
a sign difference between the definition of $K$ for phantom models (Eq. \ref{Kdef})
and for quintessence models \cite{DuttaScherrer}, corresponding
to the fact that here we are dealing with a minimum in the potenial,
rather than a maximum.

Interestingly, \eqn{finalfinal} reduces
to simple polynomial forms if $K$ is a small integer.
Some of these forms are given in \cite{DuttaScherrer}. While it would require fine tuning of the potential for $K$ to be equal to a small integer, these forms are still useful in obtaining qualitative insight into the behavior of $w\(a\)$ as a function of $K$. 

\section{Comparison to exact solutions}
We now turn to comparing our analytic
expression for the evolution of $w$ to the numerically computed exact evolution for a few different models. In each case we have a perfect fluid dark matter and a phantom field $\phi$ dynamical dark energy. 

We now consider three different phantom potentials which have local minima.
The phantom analog of the PNGB model \cite{Frieman}, has a
potential given by
\begin{equation}
\label{PNGB}
V(\phi) = \rho_{\phi 0}+M^4 \[1-\cos\(\phi/f\)\],
\end{equation}
where $M$ and $f$ are constants.
Other models with a local minimum in the potential
include the Gaussian potential,
\begin{equation}
\label{Gaussian}
V(\phi) = \rho_{\phi 0}+M^4 \[1-e^{-\phi^2/\sigma^2}\],
\end{equation}
and the quadratic potential
\begin{equation}
\label{quadratic}
V(\phi) = \rho_{\phi 0}+ V_2\phi^2.
\end{equation}
where $\sigma$ and $V_2$ are constants.

We set initial conditions deep within the matter-dominated regime.
We choose $w_0 = -1.1$ so that our results will give an upper bound
on the error in our approximation for $-1 > w_0 > -1.1$.
The value of the potential at the minimum, $V\(\phi_{*}\)$,
is chosen to be equal to the energy of the cosmological constant.
The initial value of the field $\phi_i$ is taken to be slightly
displaced from its minimum $\phi_*$, and is fixed to
give $w_0=-1.1$.
The initial velocity of the field is taken to be zero.

As discussed above, our formalism applies to
phantom models for which  \eqn{SR1} is satisfied,
but \eqn{SR2} is not.
The latter clearly holds when $K\simeq1$ but is violated
as $K$ departs from $1$. For our
specific examples, we focus on the cases $K=3$ and $K=4$. For $K=3$, for all the potentials,
\begin{eqnarray*}
\[\frac{1}{V}\frac{dV}{d\phi}\]_{a\rightarrow0}^2&\simeq& O\[10^{-1}\]\\
\left|\frac{1}{V}\frac{d^2 V}{d\phi^2}\right|_{a\rightarrow0}&\simeq&O\[1\]
\end{eqnarray*}
For $K=4$, for all the potentials, 
\begin{eqnarray*}
\[\frac{1}{V}\frac{dV}{d\phi}\]_{a\rightarrow0}^2&\simeq& O\[10^{-2}\]\\
\left|\frac{1}{V}\frac{d^2 V}{d\phi^2}\right|_{a\rightarrow0}&\simeq&O\[10\]
\end{eqnarray*}

\begin{figure}
	\epsfig{file=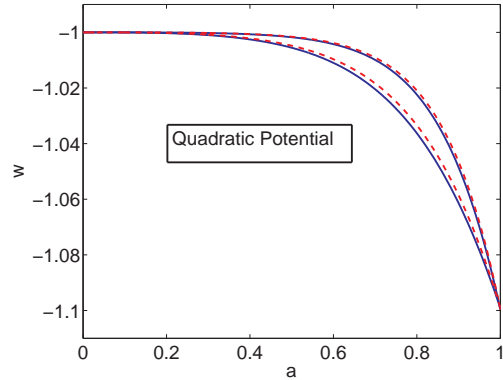,height=55mm}
	\caption
		{	\label{quad-fig} Comparison between our approximation for $w(a)$  (Eq. \ref{finalfinal}) with
	$w_0 = -1.1$ and $\Omega_{\phi 0} = 0.7$
	and the exact (numerically-integrated) evolution for $w(a)$ for
	the quadratic potential (\eqn{quadratic}).  Red (dashed)
	curves give our approximation, and solid (blue) curves
	give exact evolution, for (left to right),
	$K = 3, 4$, where $K$ is defined
	by Eq. \eqref{Kdef}.}
\end{figure}

\begin{figure}
	\epsfig{file=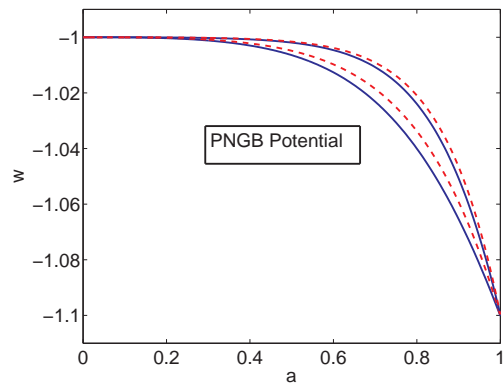,height=55mm}
	\caption
	{	\label{PNGB-fig} As Fig. 1, for the PNGB potential (\eqn{PNGB}).}
\end{figure}

\begin{figure}
	\epsfig{file=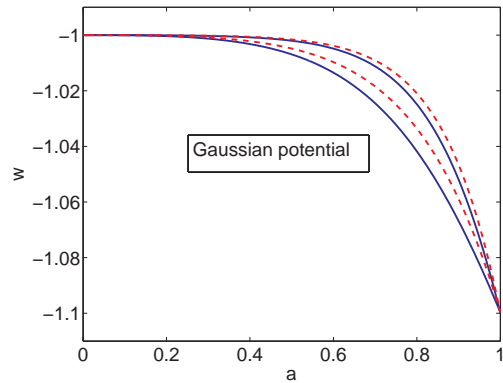,height=55mm}
	\caption
	{	\label{gaussian-fig}As Fig. 1, for the Gaussian potential (\eqn{Gaussian}).}
\end{figure}

In Figs. (\ref{quad-fig}-\ref{gaussian-fig}), the evolution of $w$ from \eqn{finalfinal}
is shown in comparison to the exact evolution for the three
different models.  The agreement, in all three cases, between Eq. (\ref{finalfinal}) and
the exact numerical evolution is excellent, with errors $\delta w \alt 0.2\%$
for the quadratic potential, $\delta w \alt 0.6\%$ for the
PNGB potential, and $\delta w \alt 0.8\%$ for the
Gaussian potential. 

We note that the accuracy is larger for the quadratic potential than it is for the other two potentials. This is
expected, since the derivation of \eqn{finalfinal}
was based on the quadratic potential.
This is however not the case for hilltop
quintessence \cite{DuttaScherrer},
where in spite of a similar derivation,
the highest accuracy was for the PNGB potential.
This is most likely due to an accidental cancellation of errors.

\begin{figure}
	\epsfig{file=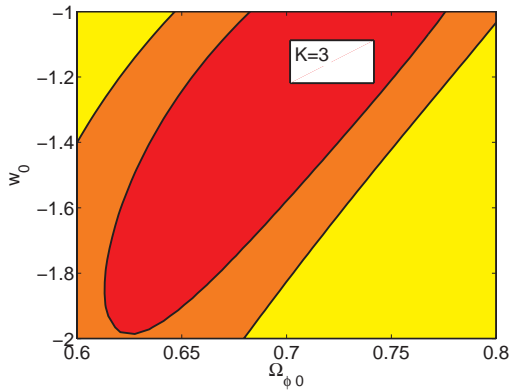,height=55mm}
	\caption
	{	\label{K3-fig}Likelihood plot from SNIa data for the parameters $w_0$ and $\Omega_{\phi0}$,
for phantom models with generic behavior
described by Eq. \eqref{finalfinal}, with $K=3$,
where $K$
is the function of the curvature of the potential
at its maximum given in Eq. \eqref{Kdef}.
The yellow (light) region
is excluded at the 2$\sigma$ level, and the darker (orange) region
is excluded at the 1$\sigma$ level.  Red (darkest) region is
not excluded at either confidence level.}
\end{figure}

\begin{figure}
	\epsfig{file=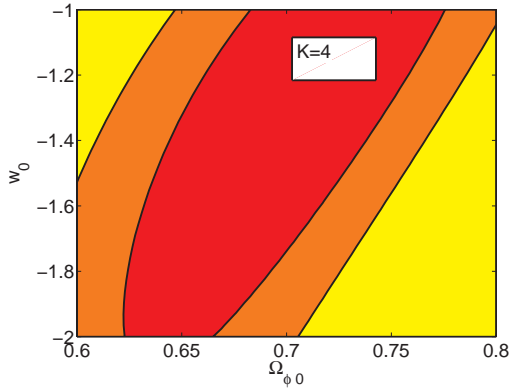,height=55mm}
	\caption
	{\label{K4-fig}
As Fig. 4, for $K=4$.}
\end{figure}

Finally, in Figs. \ref{K3-fig} and \ref{K4-fig}, we use our approximation
(Eq. \eqref{finalfinal})
to construct a $\chi^2$ likelihood plot for $w_0$
and $\Omega_{\phi0}$ with $K = 3, 4$,
using the recent Type Ia Supernovae
standard candle data (ESSENCE+SNLS+HST from \cite{Davis}).
Clearly, these models are not ruled out by current supernova data. As in the case of hilltop quintessence \cite{DuttaScherrer}, we find that a larger $K$ increases the size of the allowed region.

\eqn{finalfinal} therefore allows one to map the $w(a)$ behavior of a wide range of phantom dark energy models on to a common model-independent evolution. Clearly, this result is not well described by popular linear (CPL) parameterization of $w$ \cite{Chevallier:2000qy, Linder:2002et}, i.e., $w(a)=w_0+w_1(1-a)$, except in the special case of $K\rightarrow 1$. However, since  our treatment demonstrates that the evolution of $w$ is completely generic to any model satisfying \eqn{SR1}, and depends only on $\Omega_{\phi 0}$, $w_0$ and the curvature of the potential at its minimimum ($K$), \eqn{finalfinal} allows for an excellent parameterization of these models in terms of  $\Omega_{\phi 0}$, $w_0$ and $K$. Such a parameterization is particularly interesting in the light of recent work implying that the CPL parameterization does not satisfactorily fit the SN+BAO+CMB data simultaneously at both low and high redshifts \cite{Shafieloo:2009ti}. 

Subsequent investigations have shown that \eqn{finalfinal} has a much wider applicability than the models described in this work and \cite{DuttaScherrer}. It has been found to describe quintessence (phantom) models in which the field rolls into a minumum (maximum) of the potential, leading to a richer set of behaviors including oscillatory solutions \cite{Dutta:2009yb}. In \cite{Chiba} it was shown that \eqn{finalfinal} is not limited to models evolving near extrema of their potentials and can be applied to a broader class of slow-roll quintessence models. \cite{Chiba:2009nh} showed that \eqn{finalfinal} also applies to some slow-roll k-essence models.

\section{Conclusion}
Using techniques previously applied to quintessence,
we have derived a general expression for the evolution of $w$,
which is valid for a wide class of phantom dark energy models
in which is the field is rolling close to a local minimum.
Such models provide a mechanism to produce a value of $w$ that is
slightly less than $-1$.
We have tested our expression against the (numerically determined) exact
evolution for three different models and in each case it replicated the exact
evolution studied with an accuracy greater than $1\%$.
A comparison between our generic approximation and the observational data
indicates that
these models are allowed by SNIa data, 
and that the size of the allowed region increases
with the curvature ($V''/V$) of the potential at the minimum.

\bigskip
\bigskip
\bigskip

\acknowledgments
We thank Emmanuel Saridakis for pointing out a typo in one of our equations. R.J.S. was supported in part by the Department of Energy (DE-FG05-85ER40226). S.D. acknowledges the hospitality of the Institute for Thoeretical Science, University of Oregon, where part of this work was completed.

\end{document}